\documentclass[letterpaper]{sfchem}
\usepackage{graphicx}

\def\hh{H$_2$}
\def\nn{N$_2$}
\def\hhh{H$_3^+$}
\def\hhdp{H$_2$D$^+$}
\def\nnhp{N$_2$H$^+$}

\def\ammo{NH$_3$}
\def\dammo{ND$_3$}
\def\nhhd{NH$_2$D}
\def\nddh{ND$_2$H}
\def\hhco{H$_2$CO}
\def\chhhoh{CH$_3$OH}
\def\hho{H$_2$O}

\def\ltsim{{_<\atop{^\sim}}}
\def\kms{km~s$^{-1}$}

\def\scm{cm$^{-2}$}
\def\ccm{cm$^{-3}$}

\def\txc{$T_{\rm ex}$}

\def\mic{$\mu$m}

\begin{document}

\title{Interstellar Triply Deuterated Ammonia}

\author{Floris F.S.\ van der Tak\inst{1} \and Dariusz C.\ Lis\inst{2} \and Maryvonne
    Gerin\inst{3,4} \and Evelyne Roueff\inst{4} \and Peter
    Schilke\inst{1} \and Tom G.\ Phillips\inst{2}}

\institute{
Max-Planck-Institut f\"ur Radioastronomie, Bonn, Germany \and
California Institute of Technology, Pasadena, USA \and
Ecole Normale Sup\'erieure, Paris, France \and
Observatoire de Paris, Meudon, France} 

\authorrunning{van der Tak et al.}
\titlerunning{Interstellar Triply Deuterated Ammonia}

\maketitle 

\begin{abstract}
  
  Research on interstellar \dammo\ is reviewed and updated from the
  discovery papers. Results from observations of a dozen sources at
  centimeter and submillimeter wavelengths are presented. The two data
  sets are consistent, but do not constrain the excitation conditions
  in the \dammo-emitting gas. The column density ratios of
  \ammo/\nhhd, \nhhd/\nddh\ and \dammo/\nddh, observed in similar
  sized beams, are $\approx$10, and present a challenge for both the
  gas-phase and the grain-surface chemistry scenarios of deuterium
  fractionation.  The role of shocks and of CO depletion in deuterium
  chemistry is discussed.


\end{abstract}

\section{Introduction}
\label{sec:intro}

Deuterium-bearing molecules are important as probes of the very cold
phases of molecular clouds, prior to star formation. Moreover, the
isotopic composition of molecules is a valuable clue to their
formation mechanism.  In the gas phase, deuterium fractionation occurs
by reactions with \hhdp, CH$_2$D$^+$ and C$_2$HD$^+$ (Millar, this
volume).  Alternatively, reactions on the surfaces of dust grains may
enhance molecular D/H ratios. Both methods require low temperatures
($<$20~K) and high densities ($>10^5$~\ccm) such as found in dark
clouds and pre-stellar cores. The grain surface route also needs a
mechanism to return the molecules to the gas phase where they can be
observed by the means of submillimeter spectroscopy. The gas-phase
route can account for the deuteration of HCO$^+$ and N$_2$H$^+$, while
grain surface chemistry may be important for \hhco\ and \chhhoh.

Although astronomical detections of both singly and doubly deuterated
\ammo\ exist (\cite{roueff00}), the deuteration mechanism for \ammo\ 
is not clear.  Despite early claims, searches for \ammo\ ice based on
the 2.21~and 9.0~\mic\ features have not been successful
(\cite{taban02}), but the 3.47~\mic\ feature indicates abundances of
up to 7\% of \hho\ ice (\cite{dart02}). However, existence in the
solid state does not imply formation on grain surfaces, since \ammo\ 
may just passively accrete after forming in the gas phase.  Soon after
\cite*{rodg01} proposed \dammo\ as test of theories of interstellar
deuterium chemistry, searches were initiated.

The rotational energy levels of symmetric top molecules like \dammo\ 
are labeled by the total angular momentum $J$ and its projection on
the molecular symmetry axis $K$. Inversion splits these levels into
states which are symmetric (\textit{s}) and antisymmetric (\textit{a})
upon reflection in the plane of the D~atoms. Coupling of the $^{14}$N
nuclear spin $I$ with the rotational angular momentum $J$ causes
additional hyperfine splitting ($F=I+J$).  Current instrumentation
cannot resolve the even finer splitting due to D, which, owing to its
small quadrupole moment, is only $\approx$200~kHz.  A complete line
list is available from the CDMS catalog (\cite{mull01}) at the URL
{\tt http://www.cdms.de}.  Under the conditions favourable for
deuterium enhancement, the strongest lines of \dammo\ should be the
ground state rotational line ($1_0^a\to0_0^s$) at 309.9~GHz and the
(1,1) inversion line ($1_1^a\to1_1^s$) at 1589~MHz, which are both
accessible from the ground.

Because the Pauli exclusion principle only holds for fermions, certain
states are forbidden for \ammo\ but allowed in \dammo. For example,
the \dammo\ ground state rotational line also has a $1_0^s\to0_0^a$
component at 306.7~GHz, which does not exist for \ammo. However, spin
statistics make this line 10 times weaker than the $1_0^a\to0_0^s$
line, which is why it has escaped detection so far.

\section{Submillimeter observations}
\label{sec:cso}

\begin{table}[th]
  \begin{center}
    \caption{Upper limits for \dammo\ rotational line emission}
    \label{tab:cso}
    \begin{tabular}[lrrr]{lrrr}
\hline
\noalign{\smallskip}
Source & R.A.\ (1950) & Dec.\ (1950)     & rms\\
       & {\it h m s}  & $\circ$ $'$ $''$ & mK \\ \hline
\noalign{\smallskip}
OMC-2         & 05 32 58.6 &-05 11 42 & 24 \\
HH1-C         & 05 33 51.5 &-06 47 57 & 25 \\
NGC 2264 C    & 06 38 26.3 & 09 32 18 & 39 \\
NGC 2264 G    & 06 38 25.7 & 09 58 54 & 31 \\
Serpens S68N  & 18 27 15.2 & 01 14 47 & 30 \\
Serpens FIRS1 & 18 27 17.3 & 01 13 16 & 28 \\
\noalign{\smallskip}
\hline
    \end{tabular}
  \end{center}
\end{table}


Emission in the $J_K=1_0^a\to0_0^s$ line has been detected in the dark
cloud Barnard~1 (\cite{lis02:nd3}), two positions in the star-forming
region NGC 1333 (\cite{vdtak02}), and several positions in the
molecular cloud L~1689N (Roueff et al, in prep.). Table~\ref{tab:cso}
lists the results of a more extensive search, made with the Caltech
Submillimeter Observatory.  Sources were selected from \cite*{shah01}
to have high \nhhd\ column densities, but the positions used here are
from \cite*{jiji99} for Orion and Monoceros and from \cite*{mcmull00}
for Serpens. System temperatures were $\sim$500~K and integration
times 30--50~minutes on-source, using beam switching with a throw of
$240''$.  For an excitation temperature of 10~K, the noise levels
correspond to column density upper limits of 1.4--5.0 $10^{11}$~\scm.

\section{Centimeter observations}
\label{sec:eff}

Measurements of the pure inversion spectrum of \dammo\ go back to the
early days of microwave spectroscopy (\cite{nuck53}); many lines have
been measured since (\cite{fusina94}). However, only recently, the
hyperfine structure of the (1,1) line near 1589~MHz has been resolved
(\cite{vveld02}). 

\begin{table}[th]
  \begin{center}
    \caption{Upper limits for \dammo\ inversion line emission and absorption}
    \label{tab:eff}
    \begin{tabular}[lccr]{lccr}
\hline
\noalign{\smallskip}
Source & R.A.\ (1950) & Dec.\ (1950) & rms\\
       &{\it h m s}   & $^\circ$ $'$ $''$ & mK \\ \hline
\noalign{\smallskip}
\multicolumn{4}{c}{Emission targets} \\
\noalign{\smallskip}
\hline
\noalign{\smallskip}
NGC 1333  & 03 26 03.6 &+31 04 42 & 11-16 \\
Barnard 1 & 03 30 15.0 &+30 57 31 &  9-12 \\
CB 17     & 04 00 35.0 &+56 48 00 & 16-22 \\
L 1400 K  & 04 26 51.0 &+54 45 27 & 12-23 \\
L 1544    & 05 01 15.0 &+25 07 00 & 35-38 \\
NGC 2264  & 06 38 24.9 &+09 32 29 & 38-52 \\
L 134N    & 15 51 32.0 &-02 42 13 & 15-20 \\
L 1689N   & 16 29 27.6 &-24 22 08 & 30-46 \\
S68 FIR   & 18 27 17.5 &+01 13 23 &  9-12 \\
\noalign{\smallskip}
\hline
\noalign{\smallskip}
\multicolumn{4}{c}{Absorption targets} \\
\noalign{\smallskip}
\hline
\noalign{\smallskip}
W 33          & 18 11 19.5 & -17 56 40 & 36-46  \\
W 43          & 18 45 00.4 & -01 59 16 &101-120 \\ 
37.763-0.216  & 18 58 33.1 & +04 07 41 & 43-62  \\
W 49A         & 19 07 52.1 & +09 01 08 & 70-94  \\
\noalign{\smallskip}
\hline
    \end{tabular}
  \end{center}
\end{table}


In 2001-2002, we have used the 100-m Effelsberg telescope to search
for the $(J,K)$ = (1,1) 1589.006 MHz, (2,2) 1591.695 MHz, (3,3)
1599.645 MHz, and (4,4) 1612.997 MHz metastable inversion lines. The
front end was the 1.3-1.7~GHz primary focus receiver; the back end was
the 1024-channel autocorrelator. System temperatures were 18--33~K
depending on elevation. Integration times are 4--10 hours for NGC
1333, CB 17, S68 FIR, Barnard~1, L~1400K and L~134N, and 1--2 hours
for the other sources. Frequency switching with a throw of 1~MHz was
used for all observations.  The telescope has a beam size of $8'$ at
this wavelength.  Search targets for \dammo\ emission were selected to
have $N$(NH$_2$D) $>3\times 10^{13}$~\scm\ in an $80''$ beam
(\cite{shah01}).  We excluded Orion, where at 19~cm, the H~II region
blocks our view of the molecular cloud.

The top part of Table~\ref{tab:eff} reports the range of rms noise
levels obtained for the four lines. Only data unaffected by
interference with terrestrial signals are listed. For the (2,2) and
(3,3) lines, $\approx$10\% of the data were affected by interference,
but up to 50\% for the (4,4) line. In addition, OH 1612~MHz emission
along the line of sight made some of the (4,4) data unusable.
Fortunately, the (1,1) spectra are clear of interference, and a
follow-up search with Arecibo may prove successful.

The inversion lines were also searched for in absorption towards
bright Galactic H~II regions. Targets for this search were selected to
have $N$(H$_2$CO) $>$10$^{14}$~\scm\ as measured in 6~cm absorption
(\cite{wadiak88}) and to have 20~cm flux densities, taken from the NVSS
($45''$ resolution) of $>$1~Jy, excluding Sgr~B2 because of its low
declination. The noise levels obtained for these sources
(Table~\ref{tab:eff}, bottom part) are higher than for emission
sources, because the background H~II regions contribute significantly
to the system temperature.

The centimeter data limit $N$(\dammo) to $\ltsim$0.3--1.5$\times
10^{12}$ \scm, consistent with the submillimeter data. However, the
inversion lines can not only provide independent estimates of the
\dammo\ column density, they can also constrain its excitation. Column
densities derived from the $J_K=1_0^a\to0_0^s$ line are uncertain by
factors up to 2 through the excitation temperature. The lower
frequency of the $J_K=1_1^a\to1_1^s$ line implies a much lower
spontaneous decay rate and virtually ensures thermalization. However,
being beyond the capabilities of 100-m class telescopes such as
Effelsberg and the VLA (an effective 130-m telescope), its detection
requires 300-m class telescopes or larger.


\section{Abundance of \dammo}
\label{sec:hst}

\begin{figure}[t]
  \begin{center}
    
\resizebox{\hsize}{!}{\includegraphics[angle=-90]{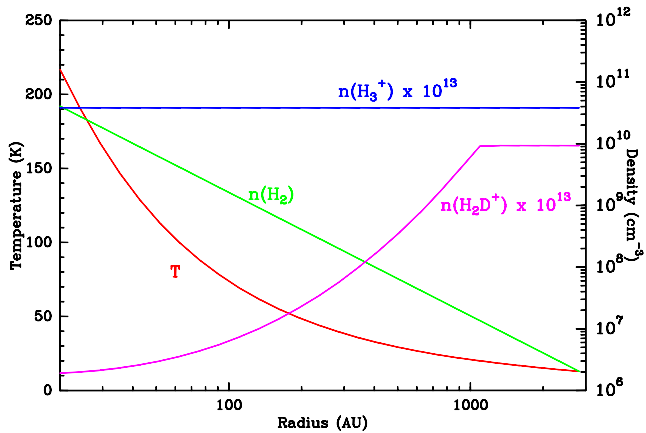}}

    \caption{Model of NGC 1333 IRAS 4A used to determine the \dammo\
    abundance. Based on van der Tak et al.\ (2002).}
    \label{fig:mdl}
  \end{center}
\end{figure}

If the excitation of \dammo\ cannot be measured, it can be calculated
through a model of the physical and chemical structure of the source.
Fig.~\ref{fig:mdl} shows the model that \cite*{vdtak02} used to
determine the \dammo\ abundance in NGC 1333 IRAS 4A. The density
structure is obtained from interferometric dust continuum
observations, while the dust temperature structure comes from a
self-consistent calculation of the thermal balance as a function of
radius. These quantities peak toward the center of the protostellar
envelope. Cosmic-ray ionization makes \hhh, at a rate that does not
vary with radius.  The reaction of \hhh\ with HD produces \hhdp, and
since it is faster at low temperatures, the \hhdp\ distribution peaks
at large radii.

The distribution of \dammo\ can either be assumed to follow that of
\hh\ or that of \hhdp. These two situations may apply to the cases of
grain-surface and gas-phase formation, respectively. The inferred
abundance of \dammo\ is $3\times 10^{-12}$ if \dammo/\hh\ is constant,
and $1\times 10^{-11}$ if \dammo/\hhdp\ is constant. Maps of the
\dammo\ emission will enable us to decide between these two
possibilities. The Sub Millimeter Array (SMA) will soon provide the
necessary resolution of $\ltsim$10$''$ at 310~GHz.

\section{Chemical implications}
\label{sec:chem}

Table~\ref{tab:abs} lists measured column densities of \ammo\
isotopo\-mers. Only in the case of NGC 1333 IRAS 4A, abundances of
\ammo\ and \dammo\ have been derived by radiative transfer modeling,
so we use column density ratios as proxies of abundance ratios. This
approach is valid as long as the excitation conditions of all
molecules are similar, which is likely true except perhaps for \ammo.

\begin{table*}[htbp]
    \caption{Observed column densities of \ammo\ isotopomers}
    \label{tab:abs}
    \begin{tabular}[lrrrrr]{lrrrrr}
\hline
\noalign{\smallskip}
Source           & $N$(\ammo)   & $N$(\nhhd)   & $N$(\nddh)   & $N$(\dammo) & \txc$^f$ \\
                 &$10^{14}$~\scm&$10^{13}$~\scm&$10^{12}$~\scm&$10^{11}$~\scm & K \\ \hline
\noalign{\smallskip}
Barnard 1        & 21$^c$       & 58$^b$      &  48$^b$      & 13$^d$  & 5 \\ 
NGC 1333 IRAS 4A & 13.8$^a$     & 39.0$^a$    & $<$4.6$^b$   & 2.9$^e$ & 10 \\ 
NGC 1333 D-peak  & 19.2$^a$     & 37.6$^a$    & $<$1.9$^b$   & 5.9$^e$ & 10 \\ \hline
\noalign{\smallskip}
Beam size ($''$) & 37           & 22           & 22           & 25  \\ 
\noalign{\smallskip}
\hline
\noalign{\smallskip}

    \end{tabular}

$^a$ From \cite*{jenny02}

$^b$ Preliminary values based on IRAM 30m observations (Gerin et al.,
in prep.)

$^c$ From Effelsberg observations

$^d$ Data from \cite*{lis02:nd3}; non-LTE calculation at
     $n$(H$_2$)=7$\times 10^4$~\ccm\ and $T$=12~K 

$^e$ From \cite*{vdtak02}

$^f$ Excitation temperature used to analyze the \nhhd\ and \nddh\ data

\end{table*}
  
The column density ratios of \ammo/\nhhd, \nhhd/\nddh\ and
\dammo/\nddh\ are each $\approx$10 (Table~\ref{tab:abs}).
This strong fractionation presents a challenge to grain
surface chemistry, and requires an atomic D/H ratio in the gas phase of
$\approx$0.15, which is $>$10$\times$ higher than in the models by
\cite*{rob00}. Unfortunately, observations do not constrain the atomic
D/H ratio in molecular clouds well. While H~I 21~cm data indicate
H/H$_2$ $\sim 10^{-3}$ (\cite{dili02}), observing D~I 92~cm is very
difficult. To determine the deuterium budget in dense interstellar
clouds, measurements of the HD 119~\mic\ line, with HIFI on Herschel
and GREAT on Sofia, will be crucial (see Phillips, this volume).

However, standard models of gas phase chemistry also have some
difficulty explaining the results of Table~\ref{tab:abs}. To reproduce
the observed column density ratios, the models require deuteron
transfer reactions to be much faster than proton transfers. Indeed,
laboratory data indicate that in the dissociative recombination of
partially deuterated ions, H is easier ejected than D
(\cite{lepetit02}). On the other hand, recent experiments on the \hhh\
+ HD reaction indicate a lower formation rate of \hhdp\ than the value
used in most chemical models (\cite{gerlich02}).


One potential problem with applying the surface chemistry route to
pre-stellar cores is their temperatures which are too low for
significant evaporation of even the most volatile ices to occur.  One
alternative may be nonthermal desorption by cosmic rays
(\cite{najita01}), but quantitative predictions do not exist yet.
However, \cite*{lis02:l1689} propose a connection between shocks and
the chemistry of deuterium. Indeed, all \dammo\ detections so far have
been at or near bipolar outflow sources. Perhaps the shocks merely
compress surrounding gas, which leads to the cold, dense conditions
favourable for deuterium chemistry. However, the shocks may also
desorb grain mantle material, while leaving the grain cores intact.
Shock velocities of $\sim$10~\kms\ would be sufficient. Such shocks
occur frequently in the interstellar medium. One possibility is that
interstellar gas is periodically shocked by ongoing star formation
activity.

Another avenue worth exploring is the connection of deuterium
chemistry with CO depletion. Freeze-out of CO, the major destroyer of
\hhdp, onto dust grains helps channeling deuterium into heavy
molecules (\cite{rob00}) and correlations between enhanced deuteration
and CO depletion have indeed been observed (\cite{jenny02,bacmann02}).
However, in the case of \ammo, such a trend does not provide direct
evidence of formation in the gas phase.  The formation of \ammo\ and
its isotopomers on grain surfaces requires a high abundance of N in
the gas phase. While in molecular clouds, most nitrogen is in \nn\ 
(\cite{womack92}), depletion of CO suppresses the formation of \nn\ 
and enhances N instead (\cite{charnley02}).  One key question is
therefore the major nitrogen carrier in pre-stellar cores. Although
neither N nor \nn\ cannot be observed directly, \nnhp\ can serve as a
tracer of \nn.  Correlation between the fractionations of \nnhp\ and
\ammo\ would thus constitute an important test of the formation
mechanism of interstellar \ammo\ and \dammo.

\begin{acknowledgements}
This paper is based on observations with the CSO and Effelsberg
telescopes. The CSO is supported by NSF grant AST 99-80846. The 100-m
telescope at Effelsberg is operated by the Max-Planck-Institut f\"ur
Radioastronomie. The authors thank Andrea Tarchi for carrying out some
of the observations presented here.
\end{acknowledgements}

\bibliographystyle{aa}
\bibliography{vdtak}

\end{document}